\documentclass[3p,twocolumn]{elsarticle}

\usepackage{hyperref}
\usepackage{graphicx}% Include figure files
\usepackage{bm}% bold math
\usepackage{textcomp,gensymb}
\usepackage[ansinew]{inputenc}

\journal{Diamond and Related Materials}

\begin{document}

\begin{frontmatter}

\title{Fabrication and structural characterization of diamond-coated tungsten tips}

\author[Laserphysik]{Alexander Tafel}\corref{mycorrespondingauthor}
\cortext[mycorrespondingauthor]{Corresponding author}
\ead{alexander.tafel@fau.de}

\author[CENEM]{Mingjian Wu}

\author[CENEM]{Erdmann Spiecker}

\author[Laserphysik]{Peter Hommelhoff}

\author[Laserphysik]{J\"urgen Ristein}

\address[Laserphysik]{Department of Physics, Friedrich-Alexander-Universit\"at Erlangen-N\"urnberg, Staudtstrasse 1, D-91058 Erlangen, Germany}
\address[CENEM]{Institute of Micro- and Nanostructure Research \& Center for Nanoanalysis and Electron Microscopy (CENEM), Department of Materials Science, Friedrich-Alexander-Universit\"at Erlangen-N\"urnberg, Cauerstrasse 6, D-91058 Erlangen, Germany}

\begin{abstract}
Coating metal nanotips with a negative electron affinity material like hydrogen-terminated diamond bears promise for a high brightness photocathode. We report a recipe on the fabrication of diamond coated tungsten tips. A tungsten wire is etched electrochemically to a nanometer sharp tip, dip-seeded in diamond suspension and subsequently overgrown with a diamond film by plasma-enhanced chemical vapor deposition. With dip-seeding only, the seeding density declines towards the tip apex due to seed migration during solvent evaporation. The migration of seeds can be counteracted by nitrogen gas flow towards the apex, which makes coating of the apex with nanometer-thin diamond possible. At moderate gas flow, diamond grows homogeneously at shaft and apex whereas at high flow diamond grows in the apex region only. With this technique, we achieve a thickness of a few tens of nanometers of diamond coating within less than 1 $\mu$m away from the apex. Conventional transmission electron microscopy (TEM), electron diffraction and electron energy loss spectroscopy confirm that the coating is composed of dense nanocrystalline diamond with a typical grain size of 20~nm. High resolution TEM reveals graphitic paths between the diamond grains.
\end{abstract}

\begin{keyword}
conformal coating \sep nanocrystalline diamond \sep EELS \sep HRTEM \sep nanoemitter
\end{keyword}

\end{frontmatter}

%\linenumbers

\section{Introduction}	
Negative electron affinity (NEA) materials are of great interest for photocathodes due to their high photoelectron yield and low thermal emittance \citep{Spicer_1977,Machuca_2000, Cui_2000}. The electron affinity of diamond depends on the exact surface, i.e. chemical species, orientation and reconstruction, and is readily adjustable between +1.7 eV and -1.3 eV with only oxygen and hydrogen as chemisorbed atoms \citep{Maier_2001}. If the surfaces are terminated by hydrogen, they reveal a comparatively low work function and true NEA, i.e. a conduction band minimum (CBM) above the vacuum level at the solid-vacuum interface. This boosts the photoelectron yield by orders of magnitude, paving the way for a highly efficient photocathode. Photo-excitation happens in the bulk and electrons are emitted into vacuum when they reach the surface, even if they have thermalized to the CBM. This is in contrast to metals, where only photoelectrons excited within the thermalization length below the surface can escape into vacuum. The combination of NEA, high thermal conductivity and mechanical robustness under imperfect vacuum condition make diamond a desirable material for photocathodes.\\
As the electron emittance - a measure of beam quality - is directly connected to the electron source size, nanosized emitters are favoured. Sharp tungsten tips are known as the brightest electron sources in scanning and transmission electron microscopes because of their extremely small virtual source size, which can be even smaller than the nanometer-sized geometrical source size \citep{Spence2013,Ehberger2015}. Coating such a sharp tungsten tip with a NEA material like hydrogen-terminated diamond holds promise for an ever brighter photocathode. Since a small source size is important to be maintained, the diamond layer should be thin. The thickness also defines the mean migration time of the excited carriers to the surface. This migration influences the electron pulse duration after pulsed photo-excitation.\\
All these arguments favour a thin and dense diamond coating on a sharp tungsten tip. For the deposition of such thin layers, a high nucleation/seeding density is crucial. The shorter the mean distance between neighbouring nucleation sites, the thinner the resulting dense film can be. Sufficiently high seed densities require appropriate adhesion of the seeds to the substrate, e.g. via electrostatic forces \cite{Hees_2011,Mandal2017}. Rheological forces occurring during the drying process of dispersed particle suspensions influence the local seeding density and can lead to phenomena like ring stains, also known as "coffee ring effect" \citep{Deegan1997}. Therefore, zeta potential adjustment and counteracting of rheological forces is essential for the control of the seeding densities. Moreover, the morphology of the nanodiamond films will play a crucial role for the electron emission properties as well. $sp^2$-bonded carbon specifically located at the grain boundaries of the film is beneficial in terms of providing sufficient conductivity through the film to prevent charging during operation. \\
Previous work reported the coating of tips based on electrophoresis \citep{Zhirnov1996, Alimova1999, Choi1996}, with bias-enhanced nucleation during chemical vapour deposition (CVD) \citep{Liu1994,Albin1999}, parafin wax and CVD \citep{Palomino2014} and ultrasonic seeding in nanodiamond slurry with and without a carburization step \citep{Tzeng2008}.
In this work, we report a new recipe on the fabrication of diamond-coated tungsten tips via dip-seeding, nitrogen gas flow and CVD. Furthermore, we characterize the structure of the resulting diamond coating. The results are promising for a high brightness diamond-based electron source. This approach is also valuable for the fabrication of samples for local electrode atom probe tomography to investigate the spatial distribution of dopants in ultrathin nanocrystalline diamond films.
\begin{figure*}
\centering
\includegraphics[width=12cm]{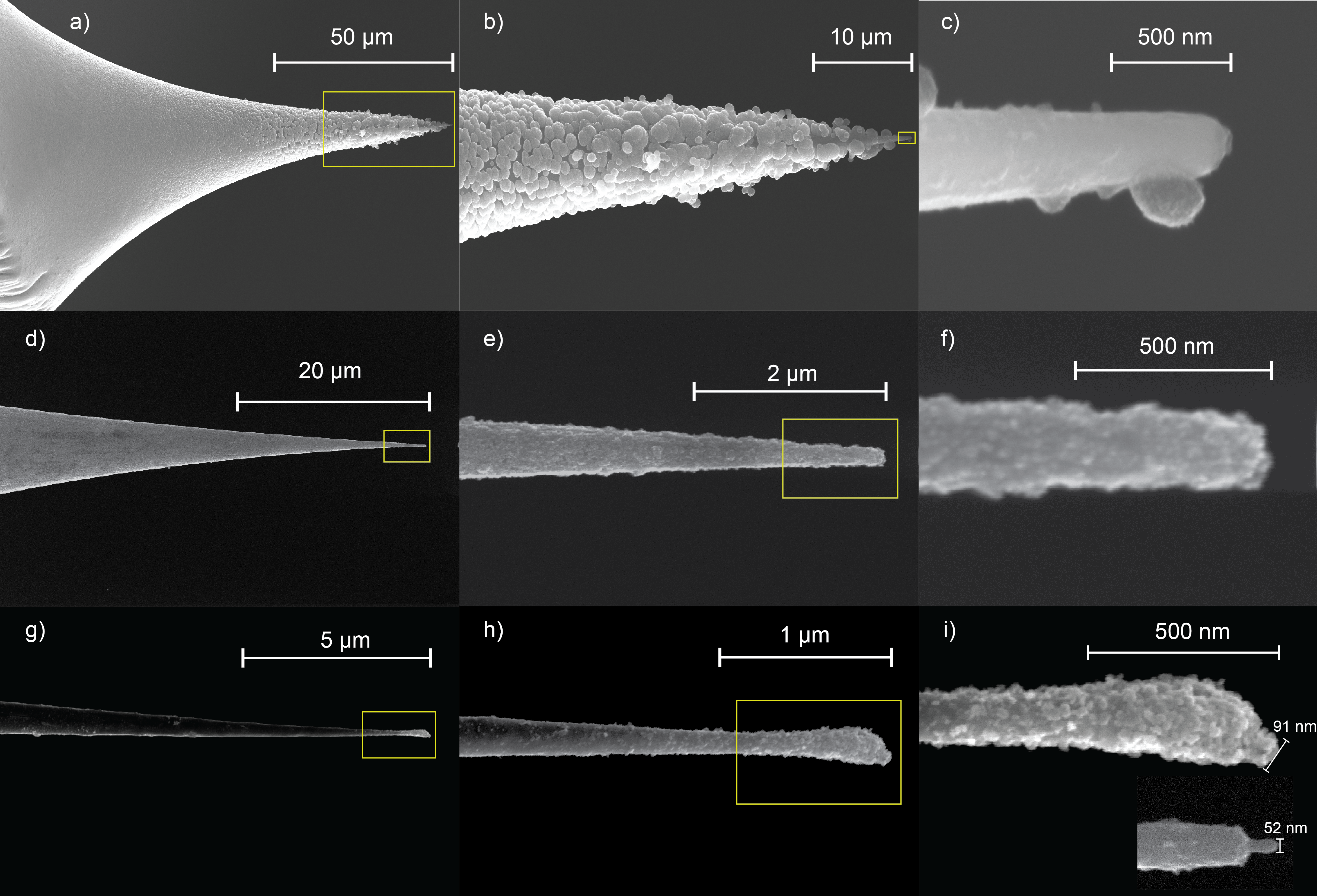}
\caption{SEM images of diamond coated tungsten tips with different nitrogen flow after dip-seeding. a)-c) Zero flow: The seeding density clearly decreases along the tip shank and only few grains are found close to the apex. Homogeneous coating of the apex could not be achieved. d)-f) Moderate flow: The entire tip including the apex is densely covered with nanodiamond. g)-i) Strong flow: Here, the apex region is selectively coated with 20~nm thin diamond. Only the first 400~nm are coated with a dense layer of diamond. From $\sim$ 800~nm behind the apex, the coating is almost absent. The inset in i) is the same tip prior to deposition.}
\label{fig:N2_flow_tips}
\end{figure*}
\section{Experimental}
\subsection{Tip fabrication}
Tungsten wire is etched electrochemically with 3 mol/L aqueous NaOH via the lamellae drop-off technique \citep{Klein1997}: A thin film of the electrolyte is trapped in a ring-shaped gold electrode. A bias voltage of 6 V is applied between this gold cathode and tungsten wire -which acts as the anode- until the wire is etched through. A second gold electrode with trapped electrolyte underneath the cathode detects the drop-off and shuts off the etching potential within less than 1 $\mu$s to prevent post-etching and blunting of the tip. Freshly etched tips are rinsed with deionised water to remove electrolyte residue. Resulting radii of tungsten tips with this method are typically 5-20~nm.
\subsection{Diamond seeding}
Sharp tungsten tips are dip-seeded for a few seconds in monodisperse nanodiamond suspensions of crystal diameter of 4-6~nm. Both 0.025 wt.\% aqueous suspension from Carbodeon and 0.025 wt.\% in dimethyl sulfoxide:methanol 1:3 from Adamas Nanotechnologies were used in this work with comparable results. Due to the oxidized tungsten surface after etching, the zeta potential of the freshly etched tungsten surface is presumably negative at the pH of the seeding suspensions. To ensure good adhesion of the diamond seeds, highly zeta-positive hydrogenated seeds are therefore used. Even though the exact zeta potential of the tungsten surface and of the seeds is not known, this qualitative approach has worked reliably on flat samples and nanotips. Experiments with zeta-negative seeds on flat tungsten samples showed more than one order of magnitude lower seeding densities. Directly after seeding of a nanotip by dip-coating, it is blown dry with pressurized nitrogen gas directed from the shaft towards the tip apex with adjustable flow rates between $0.5 - 2.5 \frac{L}{sec}$ through a nozzle with 3.5 mm diameter. 
\subsection{Diamond deposition}
Diamond deposition is performed in a home-built microwave plasma-enhanced chemical vapor deposition (MPECVD) chamber at 2.5 GHz frequency and 512 W microwave power, at a pressure of 49 mbar using 50 sccm hydrogen and 2 sccm methane flow. The sample holder is heated to 600 \degree C and then lifted into the plasma. The plasma additionally heats the sample so that the local temperature of the tip apex region is expected to be higher than 600 \degree C. Deposition times range between 2 and 20 min with a growth rate of approximately 10 $\frac{nm}{min}$. After terminating the diamond growth by switching off the microwave power and sample heater, the sample cools down in a hydrogen atmosphere at 45 mbar. This recipe reliably results in a hydrogen-terminated diamond surface with negative electron affinity \citep{Cui2000,Maier_2001}.
\subsection{Characterisation}
Scanning electron microscopy (SEM) is used for routine imaging of the tip geometry and the morphology of the diamond films. In addition, a coated tungsten tip is characterized by imaging, electron diffraction and electron energy loss spectroscopy (EELS) in a Titan Themis transmission electron microscope (TEM) operating at 200~kV. The microscope is equipped with C$_s$-correctors both at the illumination and imaging side and a Gatan GIF Quantum ER spectrometer. The sample wire was inserted in a Nanofactory STM-TEM holder. The image corrector was tuned to negative C$_s$ imaging condition where the first pass band corresponds to a resolution of 1{\AA}. We noticed that upon illumination of the tip area, the effective lens aberration can suffer from strong drift at high dose rate. Therefore, a moderate to low dose rate and careful grounding of the STM-TEM holder is necessary to obtain high-resolution TEM (HRTEM) images with good quality. The single electron energy loss (EEL) spectra were acquired directly in TEM diffraction-coupled mode with the largest collection angle - i.e. without objective aperture - to suppress the anisotropic effect in the study of graphite \citep{Leapman1983}. \\
The EEL spectrum image was acquired in scanning TEM (STEM) mode with an effective collection angle of 30 mrad, a pixel size of 1.1 nm and a short dwell time to balance the sample drift.
The low-loss and high-loss (i.e. zero-loss and Carbon 1s region in this study) are acquired with a dispersion of 0.25eV/channel. The standard Fourier-Log deconvolution method using the recorded zero-loss and plasmon peaks is applied to account for multiple scattering \citep{Egerton2011}.
We use an approximate quantification scheme to extract the $sp^2$/$sp^3$ ratio neglecting the anisotropy of the scattering cross section of the graphitic components. 
 \section{Results and discussion}
\subsection{Diamond seeding and coating}
Diamond films deposited ater dip-seeding on tungsten foils show clear signs of evaporation dynamics and their influence on local seeding density (see \ref{ch:AppendixA}). A related effect can be observed when tungsten tips are dip-seeded with nanodiamond suspensions. Without dry-blowing of the dip-seeded tips, continuous and homogeneous coating with diamond was achieved at the shank of the tip, as can be seen in fig. \ref{fig:N2_flow_tips}a)-c). The high nucleation density at the shank is presumably a result of the high positive zeta potential of seeds and the negative zeta potential of the tungsten surface. However, the density of diamond crystallites indicating the seeding density clearly decreases towards the tip. Solvent evaporation and the accompanying forces seem to redistribute the seeds, which are pushed away from the tip. To counteract this effect, we adopt a controlled flow of nitrogen gas towards the apex during the drying process. Without the nitrogen gas flow not a single sample out of ten samples was covered with diamond at the apex. \\
Using pressurized nitrogen gas for dry-blowing immediately after the dip-seeding and consecutive MPECVD, diamond was successfully grown on the tip apex with a 82\% success rate (14 out of 17 samples).  
At moderate flows rates (0.5 - 1.0 L/sec), homogeneous coating both at the shank and at the apex is achieved as can be seen in fig. \ref{fig:N2_flow_tips}d)-f). Even the sharpest tips with approximately 5~nm radius were succesfully coated with this technique (fig. \ref{fig:TEM}).
At high flow rates (up to 2.5 L/sec), the tungsten tips can even be coated selectively at the apex within less than 1 $\mu$m with 20~nm thin diamond (fig. \ref{fig:N2_flow_tips}g)-i)). From these observations we deduce that the nitrogen flow successfully counteracts the migration of the seeds away from the apex. If the flow is high enough, the seeds start migrating towards the apex and remain there only.
\subsection{Structural and chemical characterization}
\begin{figure}
\centering
\includegraphics[width=7.5 cm]{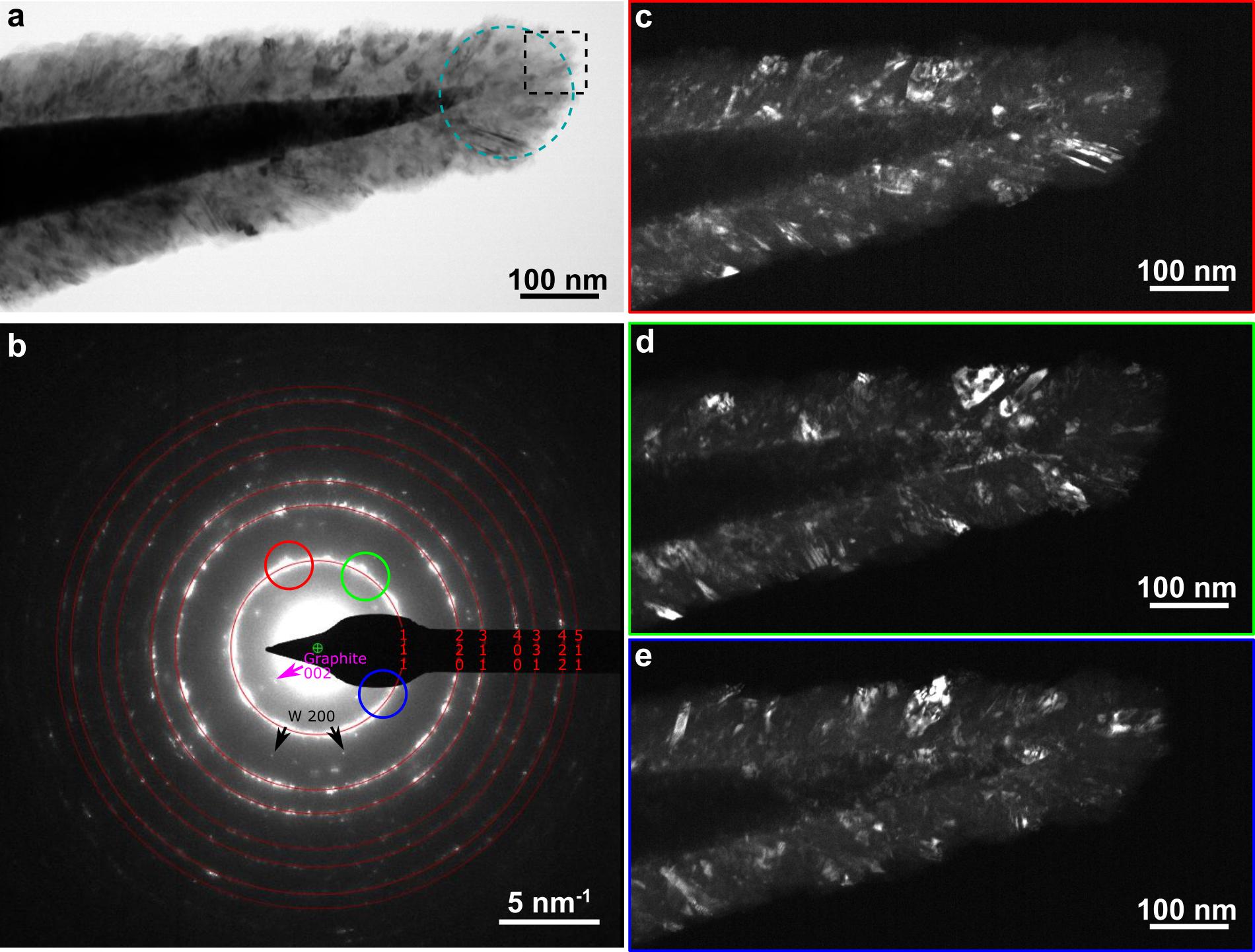}
\caption{The columnar nature of the diamond grains on the polycrystalline tungsten tip becomes visible under TEM inspection. 
(a) bright-field TEM image of the diamond coated tungsten tip. The radius of curvature of the coated tip is 100~nm, while the initial W tip radius is $\sim$ 5nm. 
(b) selected area electron diffraction (SAED) pattern of the sample. The calculated diamond powder ring pattern (using kinematical diffraction theory, red circles) is superimposed on the acquired pattern, indicating that the coating is dominated by crystalline diamond. Some weak spots can be assigned to tungsten and graphite as is exemplarily illustrated by the black arrows.
(c-e) dark-field images of the sample with the objective aperture placed at different positions of the \{111\} diffraction ring as indicated in (b) by the colored circles. Single grains oriented such that the \{111\} Bragg condition is satisfied are revealed this way and the columnar shape as well as the characteristic grain size of 20 nm is revealed. The high resolution image in  fig. \ref{fig:HRTEM} is obtained at the dashed black box in (a) and the EEL spectrum shown in  fig. \ref{fig:EELS} is acquired in the blue dashed circle region in (a).
}
\label{fig:TEM}
\end{figure}

\begin{figure}
\centering
\includegraphics[width=7.5 cm]{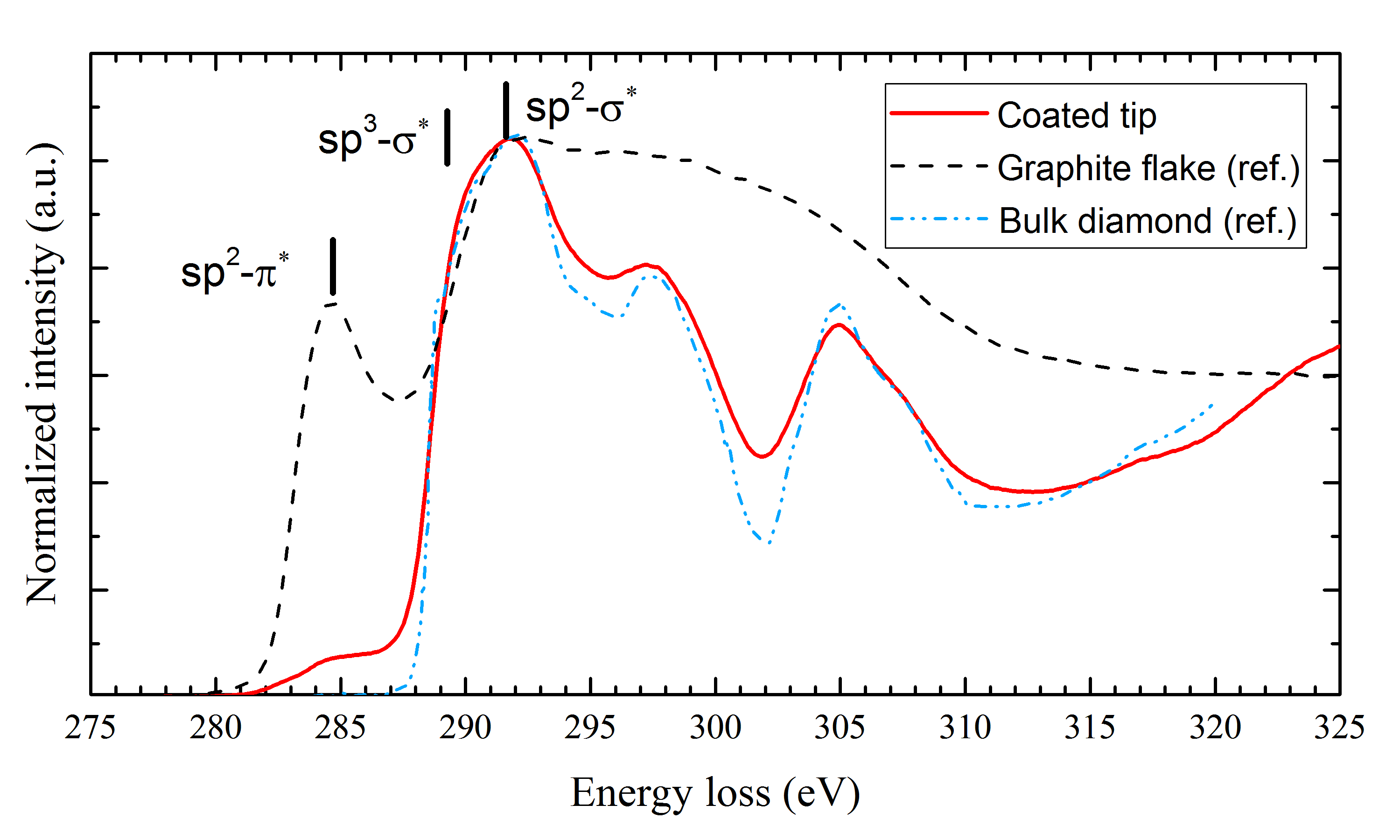}
\caption{EEL spectrum acquired in the blue dashed circle region in fig.~\ref{fig:TEM}(a) with characteristic peaks of nanocrystalline diamond, in-situ graphite reference (black) and bulk diamond (blue) \citep{Chang2016}.}
\label{fig:EELS}
\end{figure}

\begin{figure}
\centering
\includegraphics[width=7.5 cm]{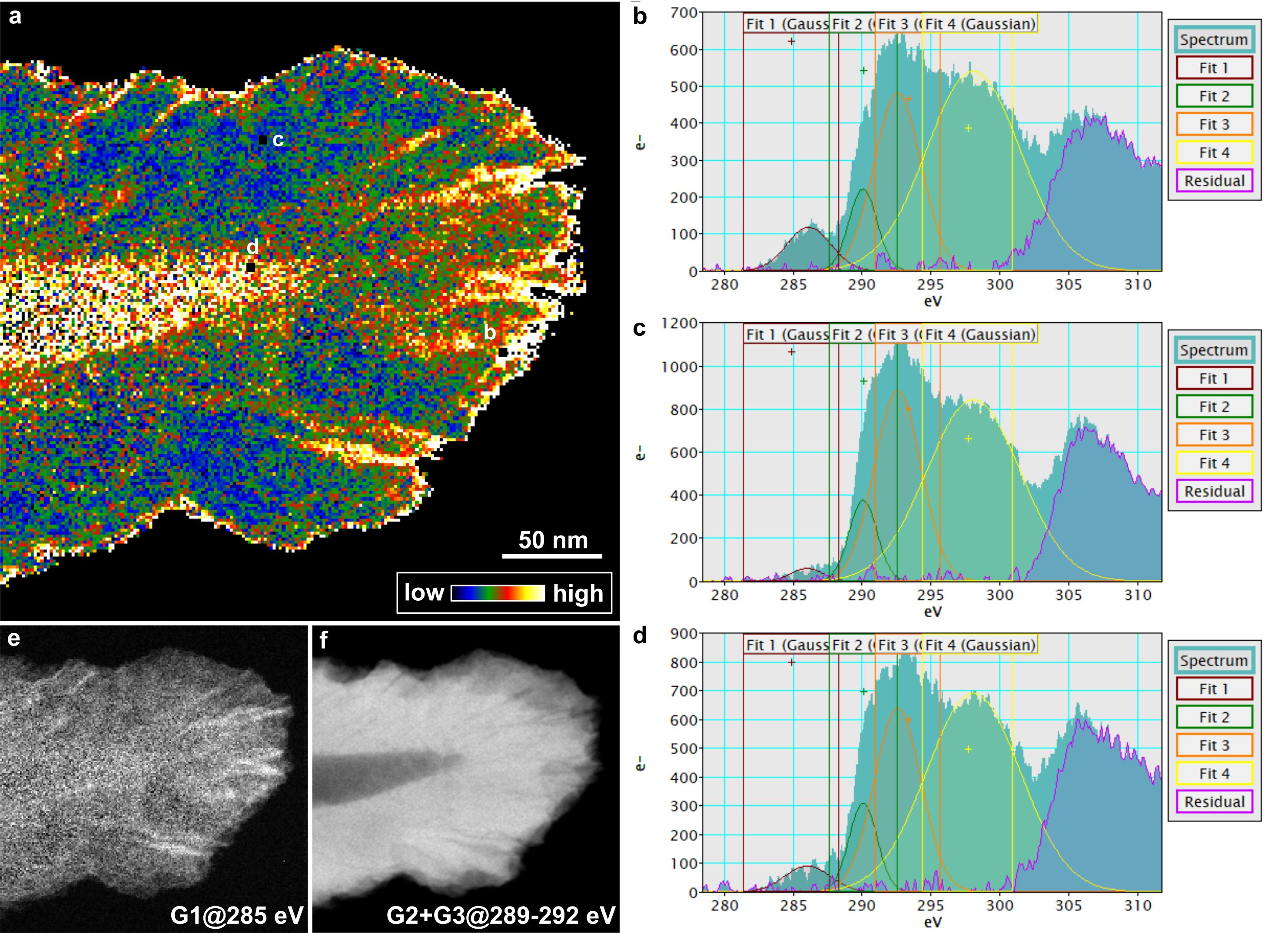}
\caption{Qualitative map of the ratio of $sp^2$- to $sp^3$-bonded carbon. (a) Evaluated map after the simple formula and processing method in \ref{ch:AppendixB}. More $sp^2$-bonded carbon is found in the axial region of the apex. The high ratio at the surface is due to carbon deposition (contamination) during the STEM-EELS measurement. The (deconvoluted and background subtracted) spectra denoted at positions b, c and d are shown in (b)-(d), respectively. (e) Intensity of extracted Gaussion peak 1 (G1) at 285~eV and (f) sum of Gaussian peak 2 and 3 (G2+G3) at 289 and 292~eV, respectively.}
\label{fig:STEM-EELS}
\end{figure}

\begin{figure}
\centering
\includegraphics[width=7.5 cm]{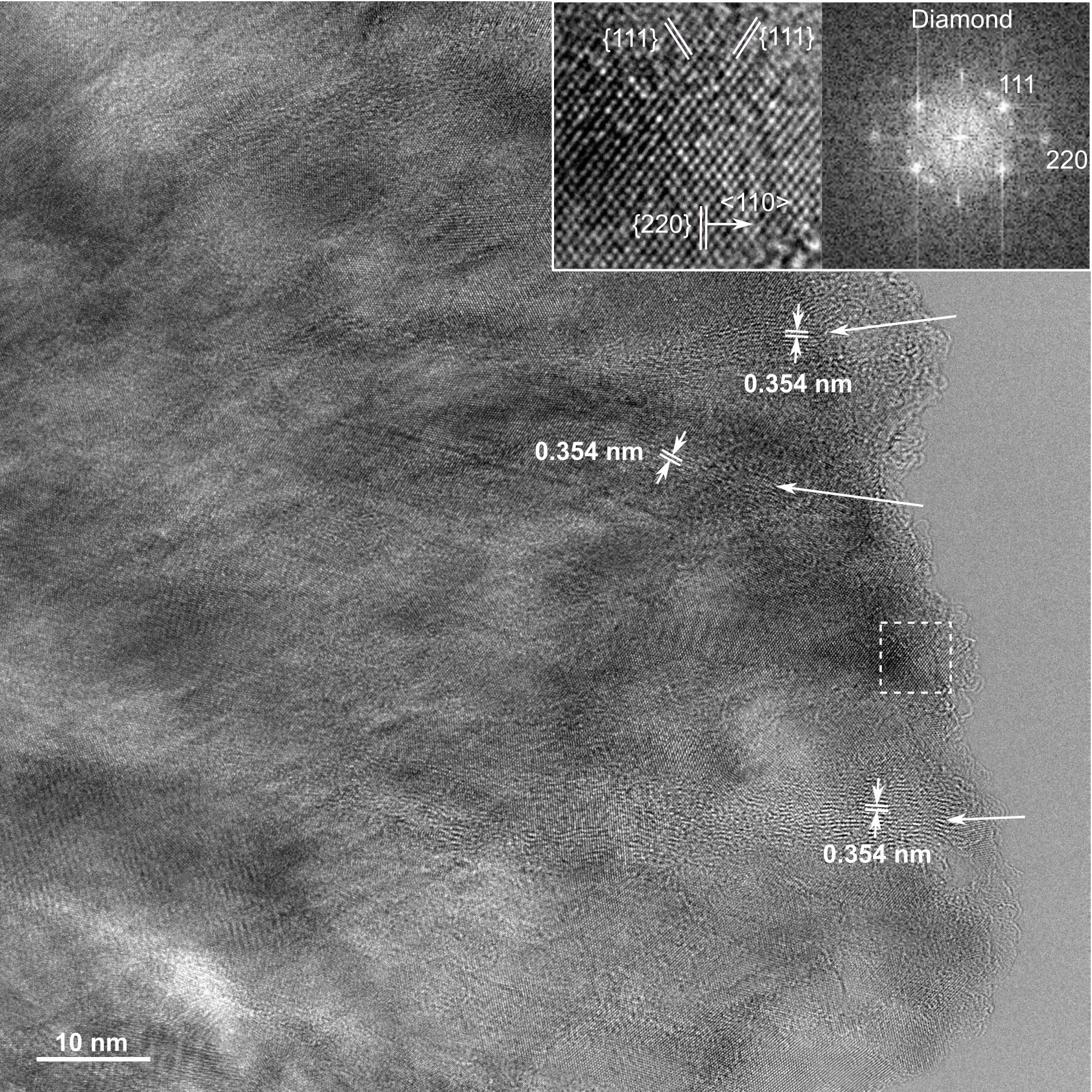}
\caption{HRTEM image of the diamond coating at the apex of the tungsten tip. Graphitic paths between the grains with interplane distance of 0.354~nm are visible.
The dashed boxed region is magnified as inset with its Fourier transform shown on the right side.}
\label{fig:HRTEM}
\end{figure}

Carbon deposited by MPECVD results in various phases as graphite, diamond and amorphous carbon depending on the exact parameters. The nanocrystalline diamond (NCD) films are expected to be composites of graphitic and diamond phases. Their morphology and composition will have decisive influence on the electron emission properties from coated tips. In order to elucidate the structural details we performed an extensive TEM study on an ultrasharp tungsten tip (apex radius approximately 5~nm) covered by a 100~nm thin NCD film.
Figure~\ref{fig:TEM}a) shows a bright-field image of the tip. The tungsten is seen dark in the center and is covered by a gray layer of NCD. The selected area electron diffraction (SAED) pattern shown in fig.~\ref{fig:TEM}b) was acquired using an aperture covering an area with about 200~nm radius over the tip region showing diffraction rings perfectly matching the powder pattern of diamond (red circles in fig. \ref{fig:TEM}b)). This confirms that the coated layer is dominated by diamond crystallites. Some additional weak spots that do not belong to diamond can be attributed to tungsten and graphite. Although the experimental diffraction ring pattern fills each circle completely, some sparse segments and strong spots can be seen especially on the \{220\} diffraction ring. \\
This is due to the textured structure of the diamond grains. However, the texture can hardly be retrieved from this diffraction pattern alone and will be subject of future research. In order to reveal the shape of the diamond grains more clearly, a series of dark-field images was recorded with the objective aperture placed at different azimuth location of the diamond \{111\} diffraction ring as indicated by the coloured circles in fig.~\ref{fig:TEM}b). The corresponding images are displayed in fig.~\ref{fig:TEM}(c-e). The columnar shape of the diamond grains with a width of about 20~nm is clearly evidenced by these dark-field images. The grain columns seem to align themselves at a small angle to the surface normal. On closer inspection, the columnar grains are also faintly visible in the bright-field image in fig.~\ref{fig:TEM}a). \\
Fig.~\ref{fig:EELS} presents the background substracted EEL spectrum recorded from the tip region marked by the dashed circle in fig.~\ref{fig:TEM}(a), as well as graphitic and diamond reference spectra. The fine structure of the carbon K-edge in the EEL spectrum reflects the orbital character of the conduction band states. Transitions into $sp^2-\sigma^*$ antibonding states, which form the upper part of the graphite conduction band, create a broad and featureless band in the EEL spectrum with a maximum at 292~eV. The diamond conduction band with $sp^3-\sigma^*$ character is also reflected as a broad band in the K-edge EEL sprectrum with a threshold at about 290 eV and peaks at 292, 297 and 305 eV. These peaks are well resolved in the spectrum of the coated tip (fig. \ref{fig:EELS}) and their presence is a clear proof that the coating is diamond \citep{Chang2016}.
Both the tip and the graphitic reference spectrum show a well resolved peak at 285 eV loss energy that is assigned to electron transitions into $sp^2-\pi^*$ antibonding states. As the $sp^2-\pi^*$ signal is absent for monocrystalline diamond, the integral of appropriate energy windows holds quantitative information on the ratio of $sp^2$- to $sp^3$-bonded carbon \citep{Lossy1995,Cuomo1991,Pappas_1992,Berger1988}. Fig. \ref{fig:STEM-EELS} shows the pixel-wise evaluated $sp^2$ to $sp^3$ ratio from the spatially-resolved STEM-EELS spectra (for details of the evalution, see \ref{ch:AppendixB}). The map reveals that the average $sp^2$ content is larger at the apex and that paths of high $sp^2$ content are present which align with the axes of the single grains (fig. \ref{fig:TEM} \& \ref{fig:STEM-EELS}). The large $sp^2$ content at the apex is attributed to a larger seeding density at the apex as fig. \ref{fig:N2_flow_tips} shows that the seeds adhere well to the apex after dry-blowing with nitrogen. 
Further insight into the morphology of the film, specifically the location of the  $sp^2$-bonded tissue, is given by the HRTEM image in fig. \ref{fig:HRTEM}. Wavy lattice fringes with a characteristic distance of 354~pm corresponding to the interlayer spacing of graphite can be seen. Some fringes are marked in white in fig. \ref{fig:HRTEM} for easier identification. Apparently, the graphitic components form contiguous paths between the diamond crystallites, which promises sufficient conductivity of the composite film to prevent charging of the tip during electron emission in future applications. \\
At the tip, one can see a few grains showing 2D lattice fringes. From the lattice fringe distances, one can attribute the lattice plane indices and plane normal directions. A small region as marked by the dotted box in fig. \ref{fig:HRTEM} is magnified in the inset with its Fourier transform on the upper right side. The \{111\} and \{220\} lattice planes can be easily recognized. However, drawing general conclusions about texture requires a thinner coating and will be subject of future studies.  
\section{Conclusion}
Tungsten tips with apex radii down to 5~nm have been successfully coated with dense nanocrystalline diamond films with a thickness as small as 20~nm. Solvent evaporation after seeding has a large effect on the local seeding density, especially at strongly curved surfaces, and must be engineered appropriately. To counteract evaporation forces, we adopt a nitrogen gas flow towards the tip apex. Diamond deposition on shaft only, apex only as well as homogeneous coating of shaft and apex is achieved by variation of the nitrogen flow. We achieved the growth of 20~nm thin diamond limited to less than 1 $\mu$m away from the tip apex by this technique. EELS and electron diffraction of a coated tungsten tip confirm the presence of diamond with a fraction of $sp^2$-bonded carbon, identified as graphitic paths inbetween grains via HRTEM images. A spatially resolved STEM-EELS measurement shows an elevated fraction of the relative $sp^2$-content at the tip apex. Furthermore, a columnar radial growth of diamond crystallites with a typical grain size of 20~nm is revealed. We expect that these diamond-coated tungsten tips with negative electron affinity offer a great potential for the use as high brightness photocathodes both in dc and ultrafast laser-triggered operation.
\section{Acknowledgements}
We thank J. Litzel for setting up and modernizing the MPECVD reactor. This work was supported by the ERC grant "Near Field Atto", DFG research training group "In-situ Microscopy with Electrons, X-rays and Scanning Probes"(GRK 1896) and DFG collaborative research center "Synthetic carbon allotropes" (CRC 953).
\section{Literature}
\bibliographystyle{prsty}
\bibliography{bibliography_with_url} % The references (bibliography) information are stored in the file named "name_of_bibliography"

\begin{thebibliography}{10}
\expandafter\ifx\csname url\endcsname\relax
  \def\url#1{\texttt{#1}}\fi
\expandafter\ifx\csname urlprefix\endcsname\relax\def\urlprefix{URL }\fi
\expandafter\ifx\csname href\endcsname\relax
  \def\href#1#2{#2} \def\path#1{#1}\fi

\bibitem{Spicer_1977}
W.~E. Spicer, Negative affinity 3{\textendash}5 photocathodes: Their physics
  and technology, Applied Physics 12~(2) (1977) 115--130.
\newblock \href {https://doi.org/10.1007/BF00896137}
  {\path{doi:10.1007/BF00896137}}.

\bibitem{Machuca_2000}
F.~Machuca, Y.~Sun, Z.~Liu, K.~Ioakeimidi, P.~Pianetta, R.~F.~W. Pease,
  Prospect for high brightness {III}{\textendash}nitride electron emitter,
  Journal of Vacuum Science {\&} Technology B: Microelectronics and Nanometer
  Structures 18~(6) (2000) 3042.
\newblock \href {https://doi.org/10.1116/1.1321270}
  {\path{doi:10.1116/1.1321270}}.

\bibitem{Cui_2000}
J.~Cui, J.~Ristein, M.~Stammler, K.~Janischowsky, G.~Kleber, L.~Ley, Hydrogen
  termination and electron emission from {CVD} diamond surfaces: a combined
  secondary electron emission, photoelectron emission microscopy, photoelectron
  yield, and field emission study, Diamond Relat. Mater. 9~(3-6) (2000)
  1143--1147.
\newblock \href {https://doi.org/10.1016/S0925-9635(99)00279-4}
  {\path{doi:10.1016/S0925-9635(99)00279-4}}.

\bibitem{Maier_2001}
F.~Maier, J.~Ristein, L.~Ley, Electron affinity of plasma-hydrogenated and
  chemically oxidized diamond (100) surfaces, Phys. Rev. B 64~(16) (2001)
  165411.
\newblock \href {https://doi.org/10.1103/PhysRevB.64.165411}
  {\path{doi:10.1103/PhysRevB.64.165411}}.

\bibitem{Spence2013}
J.~C.~H. Spence, High-Resolution Electron Microscopy, 4th Edition, Oxford
  University Press, 2013.

\bibitem{Ehberger2015}
D.~Ehberger, J.~Hammer, M.~Eisele, M.~Krüger, J.~Noe, A.~Högele, P.~Hommelhoff,
  Highly coherent electron beam from a laser-triggered tungsten needle tip,
  Phys. Rev. Lett. 114~(22) (2015) 227601.
\newblock \href {https://doi.org/10.1103/physrevlett.114.227601}
  {\path{doi:10.1103/physrevlett.114.227601}}.

\bibitem{Hees_2011}
J.~Hees, A.~Kriele, O.~A. Williams, Electrostatic self-assembly of diamond
  nanoparticles, Chem. Phys. Lett. 509~(1-3) (2011) 12--15.
\newblock \href {https://doi.org/10.1016/j.cplett.2011.04.083}
  {\path{doi:10.1016/j.cplett.2011.04.083}}.

\bibitem{Mandal2017}
S.~Mandal, E.~L.~H. Thomas, C.~Middleton, L.~Gines, J.~T. Griffiths, M.~J.
  Kappers, R.~A. Oliver, D.~J. Wallis, L.~E. Goff, S.~A. Lynch, M.~Kuball,
  O.~A. Williams, Surface zeta potential and diamond seeding on gallium nitride
  films, {ACS} Omega 2~(10) (2017) 7275--7280.
\newblock \href {https://doi.org/10.1021/acsomega.7b01069}
  {\path{doi:10.1021/acsomega.7b01069}}.

\bibitem{Deegan1997}
R.~D. Deegan, O.~Bakajin, T.~F. Dupont, G.~Huber, S.~R. Nagel, T.~A. Witten,
  Capillary flow as the cause of ring stains from dried liquid drops, Nature
  389~(6653) (1997) 827--829.
\newblock \href {https://doi.org/10.1038/39827} {\path{doi:10.1038/39827}}.

\bibitem{Zhirnov1996}
V.~Zhirnov, W.~Choi, J.~Cuomo, J.~Hren, Diamond coated si and mo field
  emitters: diamond thickness effect, Appl. Surf. Sci. 94-95 (1996) 123--128.
\newblock \href {https://doi.org/10.1016/0169-4332(95)00520-x}
  {\path{doi:10.1016/0169-4332(95)00520-x}}.

\bibitem{Alimova1999}
A.~N. Alimova, N.~N. Chubun, P.~I. Belobrov, P.~Y. Detkov, V.~V. Zhirnov,
  Electrophoresis of nanodiamond powder for cold cathode fabrication, Journal
  of Vacuum Science {\&} Technology B: Microelectronics and Nanometer
  Structures 17~(2) (1999) 715.
\newblock \href {https://doi.org/10.1116/1.590625}
  {\path{doi:10.1116/1.590625}}.

\bibitem{Choi1996}
W.~B. Choi, J.~J. Cuomo, V.~V. Zhirnov, A.~F. Myers, J.~J. Hren, Field emission
  from silicon and molybdenum tips coated with diamond powder by
  dielectrophoresis, Appl. Phys. Lett. 68~(5) (1996) 720--722.
\newblock \href {https://doi.org/10.1063/1.116585}
  {\path{doi:10.1063/1.116585}}.

\bibitem{Liu1994}
J.~Liu, V.~V. Zhirnov, G.~J. Wojak, A.~F. Myers, W.~B. Choi, J.~J. Hren, S.~D.
  Wolter, M.~T. McClure, B.~R. Stoner, J.~T. Glass, Electron emission from
  diamond coated silicon field emitters, Appl. Phys. Lett. 65~(22) (1994)
  2842--2844.
\newblock \href {https://doi.org/10.1063/1.112538}
  {\path{doi:10.1063/1.112538}}.

\bibitem{Albin1999}
S.~Albin, W.~Fu, A.~Varghese, A.~C. Lavarias, G.~R. Myneni, Diamond coated
  silicon field emitter array, Journal of Vacuum Science {\&} Technology A:
  Vacuum, Surfaces, and Films 17~(4) (1999) 2104--2108.
\newblock \href {https://doi.org/10.1116/1.581733}
  {\path{doi:10.1116/1.581733}}.

\bibitem{Palomino2014}
J.~Palomino, D.~Varshney, O.~Resto, B.~R. Weiner, G.~Morell,
  Ultrananocrystalline diamond-decorated silicon nanowire field emitters, {ACS}
  Applied Materials {\&} Interfaces 6~(16) (2014) 13815--13822.
\newblock \href {https://doi.org/10.1021/am503221t}
  {\path{doi:10.1021/am503221t}}.

\bibitem{Tzeng2008}
Y.-F. Tzeng, C.-Y. Lee, H.-T. Chiu, N.-H. Tai, I.-N. Lin, Electron field
  emission properties on ultra-nano-crystalline diamond coated silicon
  nanowires, Diamond Relat. Mater. 17~(7-10) (2008) 1817--1820.
\newblock \href {https://doi.org/10.1016/j.diamond.2008.03.023}
  {\path{doi:10.1016/j.diamond.2008.03.023}}.

\bibitem{Klein1997}
M.~Klein, G.~Schwitzgebel, An improved lamellae drop-off technique for sharp
  tip preparation in scanning tunneling microscopy, Rev. Sci. Instrum. 68~(8)
  (1997) 3099--3103.
\newblock \href {https://doi.org/10.1063/1.1148249}
  {\path{doi:10.1063/1.1148249}}.

\bibitem{Cui2000}
J.~B. Cui, M.~Stammler, J.~Ristein, L.~Ley, Role of hydrogen on field emission
  from chemical vapor deposited diamond and nanocrystalline diamond powder, J.
  Appl. Phys. 88~(6) (2000) 3667--3673.
\newblock \href {https://doi.org/10.1063/1.1288163}
  {\path{doi:10.1063/1.1288163}}.

\bibitem{Leapman1983}
R.~D. Leapman, P.~L. Fejes, J.~Silcox, Orientation dependence of core edges
  from anisotropic materials determined by inelastic scattering of fast
  electrons, Phys. Rev. B 28~(5) (1983) 2361--2373.
\newblock \href {https://doi.org/10.1103/physrevb.28.2361}
  {\path{doi:10.1103/physrevb.28.2361}}.

\bibitem{Egerton2011}
R.~Egerton, Electron Energy-Loss Spectroscopy in the Electron Microscope,
  Springer {US}, 2011.
\newblock \href {https://doi.org/10.1007/978-1-4419-9583-4}
  {\path{doi:10.1007/978-1-4419-9583-4}}.

\bibitem{Chang2016}
S.~L.~Y. Chang, A.~S. Barnard, C.~Dwyer, C.~B. Boothroyd, R.~K. Hocking,
  E.~{\={O}}sawa, R.~J. Nicholls, Counting vacancies and nitrogen-vacancy
  centers in detonation nanodiamond, Nanoscale 8~(20) (2016) 10548--10552.
\newblock \href {https://doi.org/10.1039/c6nr01888b}
  {\path{doi:10.1039/c6nr01888b}}.

\bibitem{Lossy1995}
R.~Lossy, D.~L. Pappas, R.~A. Roy, J.~P. Doyle, J.~J. Cuomo, J.~Bruley,
  Properties of amorphous diamond films prepared by a filtered cathodic arc, J.
  Appl. Phys. 77~(9) (1995) 4750--4756.
\newblock \href {https://doi.org/10.1063/1.359411}
  {\path{doi:10.1063/1.359411}}.

\bibitem{Cuomo1991}
J.~J. Cuomo, J.~P. Doyle, J.~Bruley, J.~C. Liu, Sputter deposition of dense
  diamond-like carbon films at low temperature, Appl. Phys. Lett. 58~(5) (1991)
  466--468.
\newblock \href {https://doi.org/10.1063/1.104609}
  {\path{doi:10.1063/1.104609}}.

\bibitem{Pappas_1992}
D.~L. Pappas, K.~L. Saenger, J.~Bruley, W.~Krakow, J.~J. Cuomo, T.~Gu, R.~W.
  Collins, Pulsed laser deposition of diamond-like carbon films, J. Appl. Phys.
  71~(11) (1992) 5675--5684.
\newblock \href {https://doi.org/10.1063/1.350501}
  {\path{doi:10.1063/1.350501}}.

\bibitem{Berger1988}
S.~D. Berger, D.~R. McKenzie, P.~J. Martin, {EELS} analysis of vacuum
  arc-deposited diamond-like films, Philos. Mag. Lett. 57~(6) (1988) 285--290.
\newblock \href {https://doi.org/10.1080/09500838808214715}
  {\path{doi:10.1080/09500838808214715}}.

\bibitem{Daniels2003}
H.~Daniels, A.~Brown, A.~Scott, T.~Nichells, B.~Rand, R.~Brydson, Experimental
  and theoretical evidence for the magic angle in transmission electron energy
  loss spectroscopy, Ultramicroscopy 96~(3-4) (2003) 523--534.
\newblock \href {https://doi.org/10.1016/s0304-3991(03)00113-x}
  {\path{doi:10.1016/s0304-3991(03)00113-x}}.

\end{thebibliography}
\appendix
\section{Deposition on tungsten foil} \label{ch:AppendixA}

\begin{figure}
\centering
\includegraphics[width=7.5 cm]{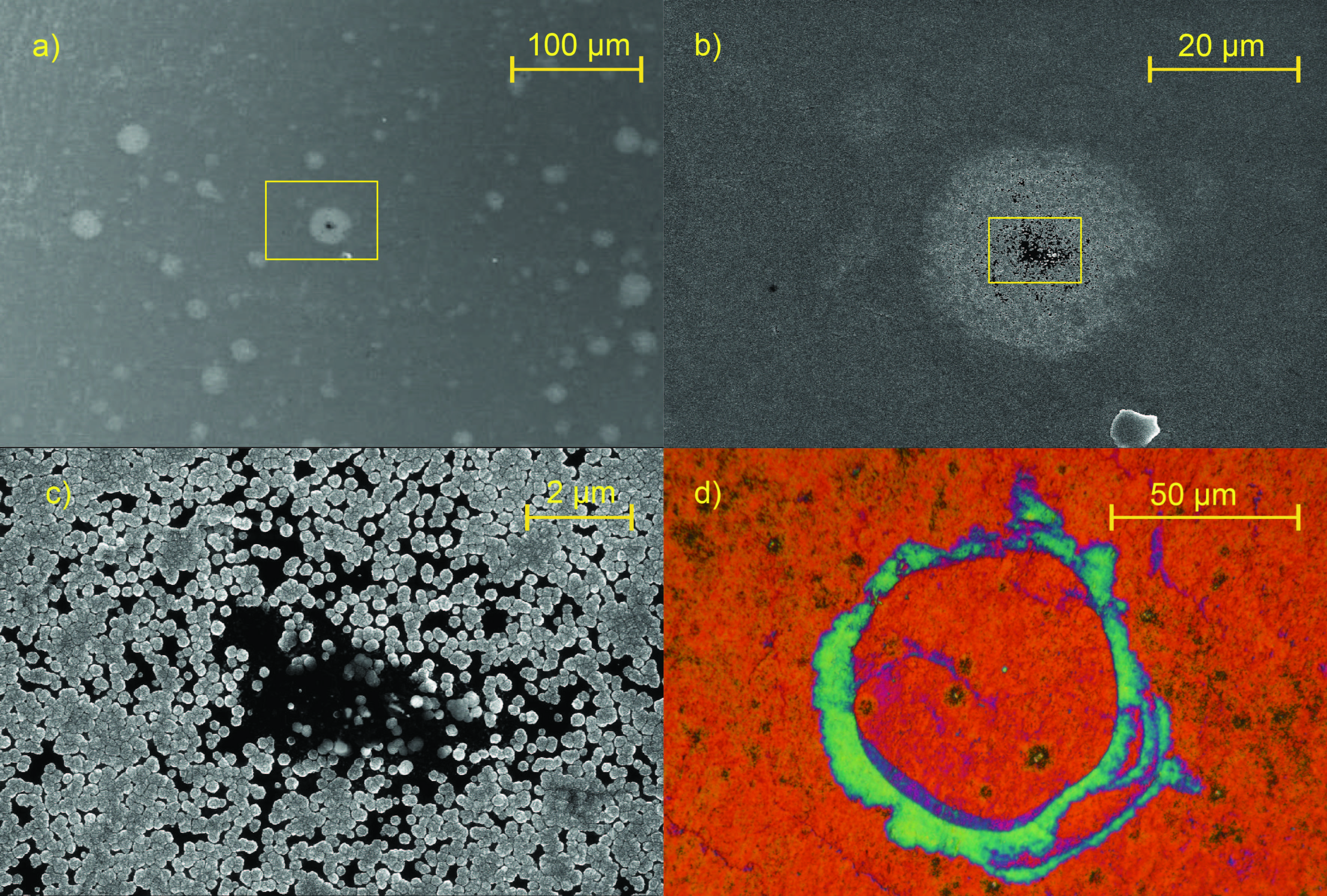}
\caption{a)-c) SEM images at various magnifications of a flat tungsten foil coated with CVD diamond after dip-seeding in nanodiamond suspension. The patterns show characteristic circular features. Seeds are pushed radially outwards during evaporation. d) Optical microscope image with circular area of larger thickness; colours arise due to thin film interference. Ring stains form due to fluid dynamics during evaporation. (Different sample than a)-c))}
\label{fig:flat_samples}

\end{figure}

Diamond deposition on dip-seeded tungsten foils have shown signs of solvent evaporation effects. Circular spots with radially growing crystallite density were often observed (fig. \ref{fig:flat_samples} a)-c), sometimes accompanyied by larger ring stains (fig. \ref{fig:flat_samples} d)). The former is a result of a evaporation front pushing seeds radially outwards whereas the latter is formed due to fluid dynamics inside the solvent happening due to evaporation \citep{Deegan1997}. These results are additional hints, that solvent evaporation plays an important role for the dip-seeding of nanotips.

\section{EELS analysis} \label{ch:AppendixB}

A mixture of $sp^2$- and $sp^3$-bonded carbon with atomic $sp^2$-fraction $x$ contains $N_{\sigma}$ $\sigma$-orbitals and $N_\pi$ $\pi$-orbitals.

\begin{equation}
	N_{\sigma} = (3 \cdot x + 4 \cdot (1-x)) \cdot N_{at}
\end{equation}
	
\begin{equation}
	N_{\pi} = x \cdot N_{at} 
\end{equation}

\begin{equation}
	\rightarrow x = \frac{4 \cdot N_\pi}{N_\sigma + N_\pi}
\end{equation}

Where $N_{at}$ is the total number of carbon atoms. Therefore the ratio of $sp^2$- to $sp^3$-bonded carbon $y$ becomes

\begin{equation}
	y = \frac{x}{1-x} = \frac{4 \cdot N_\pi}{N_\sigma - 3 \cdot N_\pi}
\end{equation}

To deconvolute the contributions in the EEL spectra, we fit three gaussians with center energy 285, 289 and 292 eV to the spectrum. Assuming that $N_\pi$ is proportional to the area under the gaussian centered at 285 eV and $N_\sigma$ is proportional to the area under the other two gaussians with the same proportionality factor, we calculate $y$ for every pixel of the STEM-EEL spectrum. This is a simplified picture of the situation, but sufficient for a spatially resolved qualitative comparison of $y$. The graphitic reference spectrum shown in fig. \ref{fig:EELS} cannot be used for a quantitative analysis due to the anisotropy of the scattering cross section with respect to crystal orientation, which was not matched between the coated tip and the reference sample. \\
Since graphite is highly anisotropic, the ratio between the inelastic scattering cross sections into $\sigma^*$ and $\pi^*$ orbitals depends strongly on the scattering angle and the angle between the incoming beam and the graphitic c-axis. Fortunately, this latter dependence vanishes for a specific so called magic scattering angle \citep{Daniels2003}. Choosing this angle in EELS experiments in combination with reference samples of known composition allows a quantitative imaging of y for arbitrary orientation of the graphitic fraction, i.e. independent of the substrate geometry and texture. Such experiments are subject of ongoing research.

\clearpage % Start a new page

\end{document}